\documentclass[runningheads]{llncs}
\usepackage[T1]{fontenc}
\usepackage[utf8]{inputenc}
\usepackage{amsmath,amssymb}
\usepackage{isabelle,isabellesym,stmaryrd}
\isabellestyle{it}
\usepackage{graphicx}

\usepackage{hyperref,color}

\newcommand{\V}{\mathbf{V}}
\newcommand{\bool}{\mathrm{bool}}

\begin{document}
\title{Wetzel: Formalisation of an Undecidable Problem Linked to the Continuum Hypothesis}
\titlerunning{Wetzel: Formalisation of an Undecidable Problem}
\author{Lawrence C Paulson\orcidID{0000-0003-0288-4279}}
\authorrunning{L C Paulson}
\institute{Computer Laboratory, University of Cambridge, UK\\
\email{lp15@cam.ac.uk} \quad
\url{https://www.cl.cam.ac.uk/~lp15/}}
%
%
%
\maketitle              
\begin{abstract}
In 1964, Paul Erd\H{o}s published a paper~\cite{erdos-interpolation} settling a question about function spaces that he had seen in a problem book.
Erd\H{o}s proved that the answer was yes if and only if the continuum hypothesis was false:
an innocent-looking question turned out to be undecidable in the axioms of ZFC\@.
The formalisation of these proofs in Isabelle/HOL demonstrate the combined use of complex analysis and set theory, and in particular how the Isabelle/HOL library for ZFC~\cite{ZFC_in_HOL-AFP} integrates set theory with higher-order logic.
\end{abstract}

\section{Introduction}

This story~\cite{garcia-wetzels-problem} expresses the richness of mathematics.
It seems that Paul Erd\H{o}s found the following question in a problem book belonging to the mathematics department of Ann Arbor University:
\begin{quote}
Suppose that $F$ is a family of analytic functions on $\mathbb{C}$ such that for each $z$ the set $\{f (z): f\in F\}$ is countable. (Call this \emph{property $P_0$}.) Then is the family $F$ itself countable?
\end{quote}
This question apparently arose in the PhD work of John E. Wetzel, in connection with spaces of harmonic functions.
Erd\H{o}s was able to show that if the continuum hypothesis failed, every family satisfying $P_0$ had to be countable; but if the hypothesis held, there existed (by a transfinite construction) an uncountable family satisfying~$P_0$.
His proof appears in \textit{Proofs from THE BOOK}~\cite{aigner-proofs},%
\footnote{End of Chapter 19. Also  \url{https://doi.org/10.1007/978-3-662-57265-8_19}}
Aigner and Ziegler's collection of ``perfect proofs'' inspired by  Erd\H{o}s.

Cantor's celebrated \textit{continuum hypothesis} (CH), number one on Hilbert's list of fundamental questions, asks whether there exists a cardinality between that of the natural numbers, namely $\aleph_0$, and that of the real numbers, namely~$\mathfrak{c}$.
Since the next cardinal after $\aleph_0$ is denoted $\aleph_1$ and the cardinality of the continuum is known to equal $2^{\aleph_0}$,  CH can be written symbolically as $\aleph_1 = 2^{\aleph_0}$.
It's fundamental because the notion of a countable set is straightforward, as is the proof by diagonalisation that the real numbers cannot be enumerated (Cantor's theorem).
CH asserts that no set exists of intermediate size between the natural numbers and the reals.
It was shown to be consistent with the axioms of set theory by Gödel and to be independent from them by Cohen.
The details can be found in any set theory text~\cite{kunen80}.

These results have been machine verified by a variety of authors.
Han and van Doorn~\cite{han-CH} were the first to mechanise Cohen's proof of the independence of CH, using Lean.%
\footnote{\url{https://leanprover.github.io}}
I have formalised Gödel's model, the \textit{constructible sets}~\cite{paulson-consistency}, and some of its properties.
Gunther et al.\ \cite{Independence_CH-AFP} have formalised model constructions both to confirm and refute CH using forcing.
These latter formalisations were done using Isabelle/ZF~\cite{paulson-gr}, the instance of Isabelle for first-order logic and set theory.
However, Isabelle/HOL is much better developed than Isabelle/ZF;
in particular, the Wetzel problem requires its complex analysis library. So this paper demonstrates how to tackle a problem that combines the worlds of analysis and set theory, including such mysteries as holomorphic functions, transfinite cardinals and recursion up to uncountable ordinals, on the common basis of higher-order logic.

The paper outlines the development of formal treatments of ZF set theory in Isabelle/HOL (\S2), continuing to the Wetzel problem in the non-CH case (\S3) and the more difficult CH case (\S4). Next comes some discussion of the work in the context of the formalisation of mathematics ganerally (\S5), and finally conclusions.

The formal development itself is available online~\cite{Wetzels_Problem-AFP} in Isabelle's Archive of Formal Proofs (AFP), as is the ZFC formalisation~\cite{ZFC_in_HOL-AFP} on which it is built.

\section{Isabelle and Set Theory}

Isabelle is a generic theorem prover, ultimately based on a uniform representation of logical syntax in the typed lambda calculus, with inference rules expressed syntactically and combined using higher-order unification~\cite{paulson-found}. Although Isabelle/HOL~\cite{isa-tutorial}---the version for higher-order logic---is by far the best known and most developed instance of Isabelle, other instances exist. These include Isabelle/FOL (classical first-order logic) and Isabelle/ZF\@.
The latter is a faithful development of set theory from the Zermelo-Fraenkel axioms within first-order logic.
Since the axiom of choice (AC) is kept separate from the other axioms, one can investigate weaker forms of AC and equivalents of AC~\cite{paulson-gr}. 

During the 1990s, I made some significant formal developments using Isabelle/ZF~\cite{paulson-reflection}, culminating in a proof of the relative consistency of the axiom of choice using Gödel's constructible universe~\cite{paulson-consistency}. Recently, several highly impressive formalisations of forcing were done within Isabelle/ZF~\cite{gunther-forcing,Independence_CH-AFP}.

On the other hand, Isabelle/ZF lacks much of the automation found in Isabelle/HOL and has no theory even of the real numbers.
That makes it unsuitable for the Wetzel problem. However, it is also possible to formalise ZF set theory within higher-order logic.

\subsection{Formalising ZF in Higher-Order Logic}

During the 1990s, Michael JC Gordon conducted several experiments~\cite{gordon-set-theory} involving the formalisation of ZF set theory in his HOL proof assistant. 
His HOL-ST simply introduced a type $\V$ of all sets and a relation ${\in}:\V\times \V\to \bool$, then asserted all the Zermelo-Fraenkel axioms. 
Sten Agerholm~\cite{agerholm-comparison} formalised Dana Scott's inverse limit construction of the set $D_\infty$, satisfying
$D_\infty \cong [D_\infty\to D_\infty]$ and yielding a model of the untyped $\lambda$-calculus.

The use of higher-order logic as opposed to first-order logic as the basis makes this version of set theory somewhat stronger than standard ZF\@. 
In HOL-ST it is possible to define the syntax of first-order logic and the semantics of set-theoretic formulas in terms of~$\V$, verifying the ZF axioms and thus proving their consistency.
On the other hand, a model for HOL-ST can be constructed in ZF plus one inaccessible cardinal. 
These remarks (which Gordon credited to Kenneth Kunen) together imply that 
the strength of HOL-ST is somewhere between ZF and ZF plus one inaccessible cardinal.
This is a weak assumption, much weaker than the dependent type theories used in Coq and Lean, which are stronger than any finite number of inaccessible cardinals~\cite{werner-sets-types}.

Some time later, Steven Obua performed a similar experiment~\cite{obua-partizan-games}, using Isabelle/HOL\@.
He adopted the same axioms and overall approach as Gordon, and demonstrated his system by formalising John H Conway's \textit{partizan games}~\cite{schleicher-conways-games}.
Obua obtained some interoperability with the existing infrastructure of Isabelle/HOL, such as its recursive function definitions. 

\subsection{Axiomatic Type Classes}

Deeper integration requires the use of axiomatic type classes, introduced by Wenzel in 1997.
An \textit{axiomatic type class}~\cite{wenzel-type} defines an open-ended collection of types on the basis of a signature and a possibly empty list of axioms. The signature specifies certain operations and their types, which will be polymorphic with type variables referring to that type class.
The operations may also be equipped with concrete syntax, such as infix declarations.
Any type, whether already existing or defined in the future, can be shown to be an instance of the class if it provides definitions of all the operations in the signature that satisfy the associated axioms.

Isabelle/HOL defines a series of type classes for orderings:
\begin{itemize}
	\item \isa{ord} introduces the relations $\le$ and $<$, of type \isa{'a\isasymRightarrow'a\isasymRightarrow bool} but no axioms
	\item \isa{preorder} extends \isa{ord} with axioms for reflexivity and transitivity
	\item \isa{order} extends \isa{preorder} with antisymmetry
	\item \isa{linorder} extends \isa{order} with the axiom \isa{x\isasymle y \isasymor\ y\isasymle x}, for linear orderings
\end{itemize}
The arithmetic types \isa{nat}, \isa{int}, \isa{real} belong to each of these classes; the type operator $\times$ preserves membership in each of these classes.

\subsection{The development ZFC-in-HOL}

My own framework~\cite{ZFC_in_HOL-AFP} for set theory within Isabelle/HOL is mathematically equivalent to Gordon's and Obua's.
Its aims were practical:
\begin{enumerate}
	\item To reproduce as much of Isabelle/ZF as possible,
	\item while achieving maximum integration with the underlying higher-order logic.
\end{enumerate}
To achieve these aims I relied on axiomatic type classes, whenever possible overloading existing symbols for set theory rather than introducing new vocabulary.

It introduces the type \isa{V} of sets and the function \isa{elts} of type \isa{V~\isasymRightarrow~V~set} mapping  a set to its elements.
Thus it uses Isabelle/HOL's existing typed sets to represent classes, but not all elements of \isa{V set} correspond to sets.
The predicate \isa{small} identifies those that are small enough, but here comes the first stage of integration: \isa{small} needs to be polymorphic, accepting a set of any type and with the more general meaning that a set is small if its elements can be put into one-to-one correspondence with the elements of a ZF set.
A set that is small in this more general sense does not in itself denote a ZF set, but this condition can be necessary if it is used in constructions that ultimately lead to a ZF set.

The fundamentals of set theory are built up as usual, starting with the ZF axioms.
Set membership is expressed using the existing (typed) set membership operator: \isa{x~\isasymin~elts y}.
Union, intersection and the subset relation are expressed using the type class \isa{distrib\_lattice} (distributive lattices), which already provides the symbols \isa{\isasymsqunion}, \isa{\isasymsqinter}, \isa{\isasymle}, etc.
Unions and intersections of families rely on the type class \isa{conditionally\_complete\_lattice}, which provides the symbols \isa{\isasymSqunion} and~\isa{\isasymSqinter}.
Type classes also allow the overloading of \isa{0} to denote the empty set (which is also the natural number zero) and \isa{1} for the natural number one.

On this foundation, it was easy to import significant chunks of Isabelle/ZF: above all, cardinal arithmetic and  the $\aleph$ operator. It was also possible to reuse Isabelle/HOL's existing theory of recursion to obtain transfinite recursion on ordinals, and hence order types, Cantor normal form and much else. So the first aim was met, but more needed to be done to achieve the second.

\subsection{The integration of ZFC-in-HOL with Isabelle/HOL}

In theory, ZFC suffices for the formalisation of all the mathematics in this problem, and much more. In practice, we absolutely do not want to be forced to develop complex analysis from first principles in set theory when it already exists in the Isabelle/HOL libraries. 
And while transfinite cardinalities and other constructions are typically understood from the framework of ZFC, they are perfectly intelligible in a broader context.

The simplest generalisation is for cardinality. The \textit{cardinality} $|x|$ of a ZF set~$x$ is simply the minimal ordinal that is equipollent to\footnote{in one-to-one correspondence with}~$x$. This definition generalises naturally to sets of any type. We can now refer to the cardinality of sets of real and complex numbers.

Next comes \textit{transfinite recursion}, also known as $\epsilon$-recursion. It is a sort of fixedpoint operator allowing the definition of functions over the whole of V\@. If a function $H$ is given, then transfinite recursion yields $F$ such that for any $a$,
\begin{align}
F(a) = H (F\restriction a), \label{eqn:transrec}
\end{align}
where $F\restriction a$ denotes $F$ itself, restricted to the elements of~$a$. It is just an instance of well-founded recursion on the membership relation. The definition is simple:
\begin{isabelle}
\isacommand{definition}\ transrec\ ::\ "((V\ \isasymRightarrow \ 'a)\ \isasymRightarrow \ V\ \isasymRightarrow \ 'a)\ \isasymRightarrow \ V\ \isasymRightarrow \ 'a"\isanewline
\ \ \isakeyword{where}\ "transrec\ H\ a\ \isasymequiv \ wfrec\ \{(x,y).\ x\ \isasymin \ elts\ y\}\ H\ a"
\end{isabelle}
This version differs from the original~\cite{ZFC_in_HOL-AFP} only in its type, which is now polymorphic as shown, allowing the recursively defined function to return anything.
Here, \isa{wfrec} is Isabelle/HOL's built-in operator for well-founded recursion.
The recursion equation~(\ref{eqn:transrec}) easily follows.
\begin{isabelle}
\isacommand{lemma}\ transrec:\ "transrec\ H\ a\ =\ H\ (\isasymlambda x\ \isasymin \ elts\ a.\ transrec\ H\ x)\ a"
\end{isabelle}
Although transfinite recursion is typically used to define operations within the set-theoretic universe~$\V$, we can now use it to create an uncountable set of analytic functions.

\subsection{Embedding Isabelle/HOL Types into $\V$}

The ZFC-in-HOL library defined the class of types that can be \textit{embedded} into the set theoretic universe, \isa{V}, by some injective map, \isa{V\_of}:
\begin{isabelle}
\isacommand{class}\ embeddable\ =\isanewline
\ \ \isakeyword{assumes}\ ex\_inj:\ "\isasymexists V\_of\ ::\ 'a\ \isasymRightarrow \ V.\ inj\ V\_of"
\end{isabelle}
If a type is embeddable then each of its elements corresponds to some ZF set.

As it happens, Isabelle/HOL already provides the class \isa{countable} of all types that can be embedded into the natural numbers. The latter are trivially embedded into \isa{V} as finite ordinals, so it is easy to show that all countable types are \isa{embeddable}.
Examples include \isa{nat}, \isa{int}, \isa{rat}, \isa{bool}.
Trivially, \isa{V} can be embedded into itself, and the type constructors $\times$, $+$ and \isa{list} are straightforwardly shown to preserve the embeddable property.

The library also defines the class of types that are small, which means that the type itself corresponds to some ZF set. It is defined in terms of the predicate \isa{small}:
\begin{isabelle}
\isacommand{class}\ small\ =\isanewline
\ \ \isakeyword{assumes}\ small:\ "small\ (UNIV::'a\ set)"
\end{isabelle}

Every \isa{countable} type (and in particular those listed above) is \isa{small}, because $\omega$ is a set. And it's obvious that every \isa{small} type is \isa{embeddable}. By proving types to be small, we further extend the embeddable class.
The type constructors $\times$, $+$ and \isa{list} preserve smallness.
The situation for the function type constructor ($\to$) is a little subtle:
\begin{isabelle}
\isacommand{instance}\ "fun"\ ::\ (small,embeddable)\ embeddable\isanewline
\isacommand{instance}\ "fun"\ ::\ (small,small)\ small\end{isabelle}
The (straightforward) proofs are omitted.
It should be obvious that we cannot expect to embed $\V\to\V$ into $\V$, but any function $f\in A\to\V$ must be a set provided $A$ is a set.

Types \isa{real} and \isa{complex} are also \isa{small}. How do we know? The reals are obtained by quotienting type \isa{nat \isasymRightarrow\ rat}. The details of their construction do not concern us.
This is the full text of the proof.
\begin{isabelle}
\isacommand{instance}\ real\ ::\ small\ \isanewline
\isacommand{proof}\ -\isanewline
\ \ \isacommand{have}\ "small\ (range\ (Rep\_real))"\isanewline
\ \ \ \ \isacommand{by}\ simp\isanewline
\ \ \isacommand{then}\ \isacommand{show}\ "OFCLASS(real,\ small\_class)"\isanewline
\ \ \ \ \isacommand{by}\ intro\_classes\isanewline
\ \ \ \ \ \ \ (metis\ Rep\_real\_inverse\ image\_inv\_f\_f\ inj\_on\_def\ replacement)\isanewline
\isacommand{qed}
\end{isabelle}
And type \isa{complex} is essentially the same thing as $\mathbb{R}^2$.

The type classes \isa{embeddable} and \isa{small} aren't especially visible in the formalisation below.
However, they provide the crucial link between type \isa{V} and other Isabelle/HOL types. 
They are the key to combining set theory and complex analysis. For example, the check that some entity is small (representable by a ZF set) is usually automatic, thanks to this type class setup.
Because \isa{real} and \isa{complex} are small, we can reason about their cardinality.

\section{Wetzel's Problem: the $\neg$CH Case}

Now it is time to formalise the proof itself, following Aigner and Ziegler's presentation~\cite{aigner-proofs}.
We begin by defining the predicate \isa{Wetzel}, corresponding to $P_0$ above, on sets of complex-valued functions. It holds if every element of the given set $F$ is analytic on the complex plane and if, for all $z$, the set $\{f(z) : f\in F\}$ is countable:
\begin{isabelle}
\isacommand{definition}\ Wetzel\ ::\ "(complex\ \isasymRightarrow \ complex)\ set\ \isasymRightarrow \ bool"\isanewline
\ \ \isakeyword{where}\ "Wetzel\ \isasymequiv \ \isasymlambda F.\ (\isasymforall f\isasymin F.\ f\ analytic\_on\ UNIV)\ \isasymand\ (\isasymforall z.\ countable((\isasymlambda f.\ f\ z)\ `\ F))"
\end{isabelle}

Remarkably, the Isabelle/HOL proof is barely 50 lines. First, here is the theorem statement:
\begin{isabelle}
\isacommand{proposition}\ Erdos\_Wetzel\_nonCH:\isanewline
\ \ \isakeyword{assumes}\ W:\ "Wetzel\ F"\ \isakeyword{and}\ NCH:\ "C\_continuum\ >\ \isasymaleph 1"\isanewline
\ \ \isakeyword{shows}\ "countable\ F"
\end{isabelle}

We set out to prove the contrapositive, negating the conclusion:
\begin{isabelle}
\ \ \isacommand{have}\ "\isasymexists z0.\ gcard\ ((\isasymlambda f.\ f\ z0)\ `\ F)\ \isasymge \ \isasymaleph 1"\ \isakeyword{if}\ "uncountable\ F"
\end{isabelle}

\noindent
Given the uncountable family $F$, we find a subset $F'\subseteq F$ of cardinality~$\aleph_1$  and thus a bijection $\phi:\aleph_1\to F'$  between the ordinals below $\aleph_1$ and $F'$.
\begin{isabelle}
\ \ \ \ \isacommand{have}\ "gcard\ F\ \isasymge \ \isasymaleph 1"\isanewline
\ \ \ \ \ \ \isacommand{using}\ that\ uncountable\_gcard\_ge\ \isacommand{by}\ force\ \isanewline
\ \ \ \ \isacommand{then}\ \isacommand{obtain}\ F'\ \isakeyword{where}\ "F'\ \isasymsubseteq \ F"\ \isakeyword{and}\ F':\ "gcard\ F'\ =\ \isasymaleph 1"\isanewline
\ \ \ \ \ \ \isacommand{by}\ (meson\ Card\_Aleph\ subset\_smaller\_gcard)\isanewline
\ \ \ \ \isacommand{then}\ \isacommand{obtain}\ \isasymphi \ \isakeyword{where}\ \isasymphi :\ "bij\_betw\ \isasymphi \ (elts\ \isasymomega 1)\ F'"\isanewline
\ \ \ \ \ \ \isacommand{by}\ (metis\ TC\_small\ eqpoll\_def\ gcard\_eqpoll)
\end{isabelle}

\noindent
We next define the family of sets $S(\alpha,\beta)$ as  $\{z.\, \phi_\alpha (z) = \phi_\beta (z)\}$.
Here $\alpha$ and $\beta$ range over ordinals, and $\omega_1$ is the first uncountable ordinal (thus the same set as $\aleph_1$, but regarded as an ordinal).
The $\phi_\alpha$ for $\alpha<\omega_1$ are the given analytic functions.
\begin{isabelle}
\ \ \ \ \isacommand{define}\ S\ \isakeyword{where}\ "S\ \isasymequiv \ \isasymlambda \isasymalpha \ \isasymbeta .\ \{z.\ \isasymphi \ \isasymalpha \ z\ =\ \isasymphi \ \isasymbeta \ z\}"
\end{isabelle}

It takes 10 lines to prove that $S(\alpha,\beta)$ is countable for $\alpha<\beta<\omega_1$,
since two distinct holomorphic functions on $\mathbb C$ can agree at only countably many arguments (a fact proved by Aigner and Ziegler).
\begin{isabelle}
\ \ \ \ \isacommand{have}\ "gcard\ (S\ \isasymalpha \ \isasymbeta )\ \isasymle \ \isasymaleph 0"\ \isakeyword{if}\ "\isasymalpha\ \isasymin\ elts\ \isasymbeta "\ "\isasymbeta\ \isasymin\ elts \isasymomega 1"\ \isakeyword{for}\ \isasymalpha \ \isasymbeta
\end{isabelle}
The next step is to define $SS$ as the union of all $S(\alpha,\beta)$ for $\alpha<\beta<\omega_1$.
\begin{isabelle}
\ \ \ \ \isacommand{define}\ SS\ \isakeyword{where}\ "SS\ \isasymequiv\ \isasymSqunion \isasymbeta \ \isasymin \ elts\ \isasymomega 1.\ \isasymSqunion \isasymalpha \ \isasymin \ elts\ \isasymbeta .\ S\ \isasymalpha \ \isasymbeta "
\end{isabelle}

A 14 line calculation shows that $|SS|\le\aleph_1$, but we are assuming the negation of CH, so $SS$ cannot be the entire complex plane: there exists some $z_0\not\in SS$.
\begin{isabelle}
\ \ \ \ \isacommand{finally}\ \isacommand{have}\ "gcard\ SS\ \isasymle \ \isasymaleph 1"\ \isacommand{.}\isanewline
\ \ \ \ \isacommand{with}\ NCH\ \isacommand{obtain}\ z0\ \isakeyword{where}\ "z0\ \isasymnotin \ SS"\isanewline
\ \ \ \ \ \ \isacommand{by}\ (metis\ Complex\_gcard\ UNIV\_eq\_I\ less\_le\_not\_le)
\end{isabelle}

That $z_0$ satisfies our requirements follows straightforwardly by the definitions of $S$ and $SS$.

\section{Wetzel's Problem: the CH Case}

Assuming CH, it is possible to construct an uncountable family of analytic functions that makes the Wetzel property $P_0$ fail.
It's a rather delicate transfinite recursion, so let's review the argument before examining the formal proof.

\subsection{The transfinite construction}

CH implies that $|\mathbb{C}| = \aleph_1$ and we can write $\mathbb{C} = \{\zeta_\alpha : \alpha < \omega_1 \}$.
Now consider the set $D\subseteq\mathbb{C}$ of \textit{rational} complex numbers: 
\begin{align*}
 D = \{p+iq: p,q\in\mathbb{Q}\}.
\end{align*}
Suppose that we had a family of functions 
 $\{f_\beta : \beta < \omega_1 \}$ such that 
\begin{align}\label{eqn:f_in_D}
	f_\beta (\zeta_\alpha) \in D\quad\text{if}\quad\alpha<\beta.
\end{align} 
 Since the set $D$ is countable and the $\zeta_\alpha$ for $\alpha < \omega_1$ include all the complex numbers, the desired result would follow.

Erd\H{o}s showed how to construct this family by transfinite induction. For an arbitrary $\gamma<\omega_1$, assume that a family of distinct analytic functions $\{f_\beta: \beta<\gamma\}$ is defined below~$\gamma$.
To conclude the inductive argument and therefore prove the theorem, we must extend it with a new function, $f_\gamma$.

Since $\gamma$ is countable, the set $\{f_\beta: \beta<\gamma\}$ can be enumerated as $\{g_0, g_1, \ldots\}$
and $\{\zeta_\alpha : \alpha < \gamma\}$ can be enumerated as $\{w_0, w_1, \ldots\}$;
both sets are finite or infinite according to whether $\gamma$ itself is finite or infinite.
The sought-for analytic function $f_\gamma$ should satisfy, for all $n$,
\begin{align}
	f_\gamma (w_n) \in D \quad\text{and}\quad f_\gamma (w_n) \not= g_n(w_n). \label{eqn:f_induct}
\end{align}
The second condition above ensures that $f_\gamma$ is new by diagonalisation, while the first is simply~(\ref{eqn:f_in_D}).
We construct $f_\gamma$ by putting
\begin{align}
\begin{split}
f_\gamma(z) & := \epsilon_0 + \epsilon_1 (z - w_0) + \epsilon_2 (z - w_0)(z - w_1)\\
	       & \quad + \epsilon_3 (z - w_0)(z - w_1)(z - w_2) + \cdots.	\label{eqn:f_gamma}
\end{split}
\end{align}
Again, this sum is finite or infinite according to $\gamma$. 
The $\{\epsilon_m\}$ are complex numbers chosen one at a time to satisfy conditions~(\ref{eqn:f_induct}) above; 
ensuring that $f_\gamma (w_n)$ avoids $g_n(w_n)$ is possible because $D$ is dense in~$\mathbb{C}$\@.
In the finite case the sum is a polynomial, so trivially analytic. 
In the infinite case, the choice of the $\{\epsilon_m\}$ needs to be carefully calibrated in order to satisfy the conditions while converging to zero sufficiently rapidly. 
We have a lot of leeway and I chose 
\begin{align}
|\epsilon_m| < \bigl[m! \cdot \prod_{i<m} (1 + |w_i|)\bigr]^{-1}.  \label{eqn:epsilon_bound}
\end{align}
The summation~(\ref{eqn:f_gamma}) converges to an analytic function because the uniform limit of holomorphic functions is holomorphic, and the limit can be shown to be uniform by the Weierstrass $M$-test.%
\footnote{Thanks to Manuel Eberl for suggesting this argument.}

A clever aspect of the construction is that the conditions~(\ref{eqn:f_induct}) constrain only $f_\gamma(w_n)$, 
whose value depends only on $\epsilon_m$ for $m\le n$. That is, $f_\gamma(w_0) = \epsilon_0$, $f_\gamma(w_1) = \epsilon_0 + \epsilon_1 (w_1 - w_0)$, etc.
The desired values of $\epsilon_m$ can be calculated sequentially.

\subsection{The Isabelle/HOL Formalisation}

\newcommand{\prf}{$\langle$proof$\,\rangle$}

The full proof is some 280 lines of Isar text. Let's examine some highlights.
First, let's state the theorem formally:
\begin{isabelle}
\isacommand{proposition}\ Erdos\_Wetzel\_CH:\isanewline
\ \ \isakeyword{assumes}\ CH:\ "C\_continuum\ =\ \isasymaleph 1"\isanewline
\ \ \isakeyword{obtains}\ F\ \isakeyword{where}\ "Wetzel\ F"\ \isakeyword{and}\ "uncountable\ F"
\end{isabelle}
The proof begins with a self-evident definition of $D$. It's then shown to be countably infinite.
\begin{isabelle}
\ \ \isacommand{define}\ D\ \isakeyword{where}\ "D\ \isasymequiv \ \{z.\ Re\ z\ \isasymin \ \isasymrat \ \isasymand \ Im\ z\ \isasymin \ \isasymrat \}"\isanewline
\ \ \isacommand{have}\ Deq:\ "D\ =\ (\isasymUnion x\isasymin \isasymrat .\ \isasymUnion y\isasymin \isasymrat .\ \{Complex\ x\ y\})"\isanewline
\ \ \ \ \isacommand{using}\ complex.collapse\ \isacommand{by}\ (force\ simp:\ D\_def)\isanewline
\ \ \isacommand{with}\ countable\_rat\ \isacommand{have}\ "countable\ D"\isanewline
\ \ \ \ \isacommand{by}\ blast\isanewline
\ \ \isacommand{have}\ "infinite\ D"\isanewline
\ \ \ \ \prf\isanewline
\ \ \isacommand{have}\ "\isasymexists w.\ Re\ w\ \isasymin \ \isasymrat \ \isasymand \ Im\ w\ \isasymin \ \isasymrat \ \isasymand \ norm\ (w\ -\ z)\ <\ e"\ \isakeyword{if}\ "e\ >\ 0"\ \isakeyword{for}\ z\ e\isanewline
\ \ \ \ \prf\isanewline
\ \ \isacommand{then}\ \isacommand{have}\ cloD:\ "closure\ D\ =\ UNIV"\isanewline
\ \ \ \ \isacommand{by}\ (auto\ simp:\ D\_def\ closure\_approachable\ dist\_complex\_def)
\end{isabelle}
The closure of $D$ equals the universal set \isa{UNIV} of type \isa{complex set}.
We obtain the transfinite enumeration $\{\zeta_\alpha : \alpha < \omega_1\}$ of the complex plane.
 \begin{isabelle}
\ \ \isacommand{obtain}\ \isasymzeta \ \isakeyword{where}\ \isasymzeta :\ "bij\_betw\ \isasymzeta \ (elts\ \isasymomega 1)\ (UNIV::complex\ set)"\isanewline
\ \ \ \ \isacommand{by}\ (metis\ Complex\_gcard\ TC\_small\ assms\ eqpoll\_def\ gcard\_eqpoll)
\end{isabelle}
Next come some technical definitions: \isa{inD} for functions whose range for certain arguments lies within $D$ and $\Phi$ to express that $f$ is a family of analytic functions indexed by the ordinals up to~$\beta$. 

\begin{isabelle}
\ \ \isacommand{define}\ inD\ \isakeyword{where}\ "inD\ \isasymequiv \ \isasymlambda \isasymbeta \ f.\ (\isasymforall \isasymalpha \ \isasymin \ elts\ \isasymbeta .\ f\ (\isasymzeta \ \isasymalpha )\ \isasymin \ D)"\isanewline
\ \ \isacommand{define}\ "\isasymPhi \ \isasymequiv\ \isasymlambda \isasymbeta \ f.\ f\ \isasymbeta \ analytic\_on\ UNIV\ \isasymand \ inD\ \isasymbeta \ (f\ \isasymbeta )\ \isasymand \ inj\_on\ f\ (elts(succ\ \isasymbeta ))"
\end{isabelle}

\subsection{The Transfinite Construction}

The following lemma is the step of the transfinite induction.
A family $f$ defined below the ordinal~$\gamma$ is extended with a new function, $f_\gamma$ (the variable $h$ below).

\begin{isabelle}
\ \ \isacommand{have}\ *:\ "\isasymexists h.\ \isasymPhi \ \isasymgamma \ ((restrict\ f\ (elts\ \isasymgamma ))(\isasymgamma :=h))"\isanewline
\ \ \ \ \isakeyword{if}\ \isasymgamma :\ "\isasymgamma \ \isasymin \ elts\ \isasymomega 1"\ \isakeyword{and}\ "\isasymforall \isasymbeta \ \isasymin \ elts\ \isasymgamma .\ \isasymPhi \ \isasymbeta \ f"\ \isakeyword{for}\ \isasymgamma \ f
\end{isabelle}

\noindent
The construction of $h$ depends on whether or not $\gamma$ is finite:
\begin{isabelle}
\ \ \ \ \isacommand{obtain}\ h\ \isakeyword{where}\ "h\ analytic\_on\ UNIV"\ \ "inD\ \isasymgamma \ h"\ \ "\isasymforall \isasymbeta \ \isasymin \ elts\ \isasymgamma .\ h\ \isasymnoteq \ f\ \isasymbeta"\isanewline
\ \ \ \ \isacommand{proof}\ (cases\ "finite\ (elts\ \isasymgamma )")
\end{isabelle}
The finite case is easier, since the function we have to construct will simply be a polynomial, and trivially analytic. The finite ordinal~$\gamma$ is simply some natural number~$n$, and the construction of $f_\gamma$ is by induction on~$n$. 

Let's examine the infinite case, in which similar ideas are taken to the max.
We formalise the step of writing $\{f_\beta: \beta<\gamma\}$ as $\{g_0, g_1, \ldots\}$
and $\{\zeta_\alpha : \alpha < \gamma\}$ as $\{w_0, w_1, \ldots\}$
by picking some bijection $\eta$ between $\mathbb{N}$ and $\gamma$, a countable ordinal which of course equals $\{\alpha:\alpha<\gamma\}$.

\begin{isabelle}
\ \ \ \ \ \ \isacommand{case}\ False\isanewline
\ \ \ \ \ \ \isacommand{then}\ \isacommand{obtain}\ \isasymeta \ \isakeyword{where}\ \isasymeta :\ "bij\_betw\ \isasymeta \ (UNIV::nat\ set)\ (elts\ \isasymgamma )"\isanewline
\ \ \ \ \ \ \ \ \isacommand{by}\ (meson\ \isasymgamma \ countable\_infiniteE'\ less\_\isasymomega 1\_imp\_countable)\isanewline
\ \ \ \ \ \ \isacommand{define}\ g\ \isakeyword{where}\ "g\ \isasymequiv \ f\ o\ \isasymeta "\isanewline
\ \ \ \ \ \ \isacommand{define}\ w\ \isakeyword{where}\ "w\ \isasymequiv \ \isasymzeta \ o\ \isasymeta "
\end{isabelle}

The following three definitions set up the finite approximants of our new analytic function, $h$.
Note that $p(n)$ is the product of $z - w_i$ for $i<n$, while $q(n)$ is $\prod_{i<n} (1 + |w_i|)$.

\begin{isabelle}
\ \ \ \ \ \ \isacommand{define}\ p\ \isakeyword{where}\ "p\ \isasymequiv \ \isasymlambda n\ z.\ \isasymProd i<n.\ z\ -\ w\ i"\isanewline
\ \ \ \ \ \ \isacommand{define}\ q\ \isakeyword{where}\ "q\ \isasymequiv \ \isasymlambda n.\ \isasymProd i<n.\ 1\ +\ norm\ (w\ i)"\isanewline
\ \ \ \ \ \ \isacommand{define}\ h\ \isakeyword{where}\ "h\ \isasymequiv \ \isasymlambda n\ \isasymepsilon \ z.\ \isasymSum i<n.\ \isasymepsilon \ i\ *\ p\ i\ z"
\end{isabelle}

The following three definitions constrain the next value~$\epsilon_n$, a calculation left unspecified by previous authors~\cite{aigner-proofs,erdos-interpolation}. To ensure convergence, each approximant needs to be close to the previous one, hence a neighbourhood (open ball) whose radius has $n!q(n)$ for its denominator. \isa{DD} is the intersection of this neighbourhood with the set~$D$, excluding the one point that must be avoided. Finally, the function \isa{dd} picks an arbitrary element of~\isa{DD}\@.
In each of these functions, the argument $\epsilon$ is an integer-valued function representing the $\{\epsilon_i\}$ for $i<n$.

\begin{isabelle}
\ \ \ \ \ \ \isacommand{define}\ BALL\ \isakeyword{where}\isanewline
\ \ \ \ \ \ \ \ \ \ \ \ \ \ "BALL\ \isasymequiv \ \isasymlambda n\ \isasymepsilon .\ ball\ (h\ n\ \isasymepsilon \ (w\ n))\ (norm\ (p\ n\ (w\ n))\ /\ (fact\ n\ *\ q\ n))"\isanewline
\ \ \ \ \ \ \isacommand{define}\ DD\ \isakeyword{where}\ "DD\ \isasymequiv \ \isasymlambda n\ \isasymepsilon .\ D\ \isasyminter \ BALL\ n\ \isasymepsilon \ -\ \{g\ n\ (w\ n)\}"\isanewline
\ \ \ \ \ \ \isacommand{define}\ dd\ \isakeyword{where}\ "dd\ \isasymequiv \ \isasymlambda n\ \isasymepsilon .\ SOME\ x.\ x\ \isasymin \ DD\ n\ \isasymepsilon"
\end{isabelle}
Because $D$ is dense in $\mathbb{C}$, the set \isa{DD} is always nonempty. Therefore, \isa{dd} always chooses an element satisfying the constraints described above.

\begin{isabelle}
\ \ \ \ \ \ \isacommand{have}\ "DD\ n\ \isasymepsilon \ \isasymnoteq \ \{\}"\ \isakeyword{for}\ n\ \isasymepsilon \isanewline
\ \ \ \ \ \ \ \ \prf\isanewline
\ \ \ \ \ \ \isacommand{then}\ \isacommand{have}\ dd\_in\_DD:\ "dd\ n\ \isasymepsilon \ \isasymin \ DD\ n\ \isasymepsilon "\ \isakeyword{for}\ n\ \isasymepsilon \isanewline
\ \ \ \ \ \ \ \ \isacommand{by}\ (simp\ add:\ dd\_def\ some\_in\_eq)
\end{isabelle}

Each $\epsilon_n$ (written  \isa{coeff~n} in Isabelle) is defined in terms of the previous epsilons by course of values recursion. 
The ugly details of the definition are omitted but the following line shows the recursion that it satisfies; note that \hbox{\isa{h\ n\ coeff\ (w\ n)}} refers recursively to all the $\epsilon_i$ for $i<n$. 
The last line verifies the bound~(\ref{eqn:epsilon_bound}).

\begin{isabelle}
\ \ \ \ \ \ \isacommand{define}\ coeff\ \isakeyword{where}\ "coeff\ \isasymequiv\ \ldots"\isanewline
\ \ \ \ \ \ \isacommand{have}\ coeff\_eq:\ "coeff\ n\ =\ (dd\ n\ coeff\ -\ h\ n\ coeff\ (w\ n))\ /\ p\ n\ (w\ n)"\ \isakeyword{for}\ n\isanewline
\ \ \ \ \ \ \ \ \isacommand{by}\ (simp\ add:\ def\_wfrec\ [OF\ coeff\_def])\isanewline
\ \ \ \ \ \ \isacommand{have}\ norm\_coeff:\ "norm\ (coeff\ n)\ <\ 1\ /\ (fact\ n\ *\ q\ n)"\ \isakeyword{for}\ n\isanewline
\ \ \ \ \ \ \ \ \ \prf
\end{isabelle}

To conclude, define the function~$f_\gamma$ (calling it \isa{hh}) and prove that it is analytic. 

\begin{isabelle}
\ \ \ \ \ \ \isacommand{define}\ hh\ \isakeyword{where}\ "hh\ \isasymequiv \ \isasymlambda z.\ suminf\ (\isasymlambda i.\ coeff\ i\ *\ p\ i\ z)"\isanewline
\ \ \ \ \ \ \isacommand{have}\ "hh\ holomorphic\_on\ UNIV"\isanewline
\ \ \ \ \ \ \ \ \prf\isanewline
\ \ \ \ \ \ \isacommand{then}\ \isacommand{show}\ "hh\ analytic\_on\ UNIV"\isanewline
\ \ \ \ \ \ \ \ \isacommand{by}\ (simp\ add:\ analytic\_on\_open)
\end{isabelle}

Here are a few details of the proof that was skipped above.
For a fixed $n$ and complex number~$z$, we show uniform convergence within a circle of radius~1.
First, we need a bound on $\prod_{i<n} |z' - w_i|$ for $|z-z'|< 1$:

\begin{isabelle}
\ \ \ \ \ \ \isacommand{have}\ norm\_p\_bound:\ "norm\ (p\ n\ z')\ \isasymle \ q\ n\ *\ (1\ +\ norm\ z)\ \isacharcircum \ n"\ \isanewline
\ \ \ \ \ \ \ \ \ \ \isakeyword{if}\ "dist\ z\ z'\ \isasymle \ 1"\ \isakeyword{for}\ n\ z\ z'\isanewline
\ \ \ \ \ \ \ \prf
\end{isabelle}

To apply the Weierstrass $M$-test to~\isa{hh}, exhibit a summable series $M$ whose terms bound the finite approximants of the summation, which are given by~$h$:
\begin{isabelle}
\ \ \ \ \ \ \isacommand{have}\ "uniform\_limit\ (cball\ z\ 1)\ (\isasymlambda n.\ h\ n\ coeff)\ hh\ sequentially"\isanewline
\ \ \ \ \ \ \ \ \isacommand{unfolding}\ hh\_def\ h\_def\isanewline
\ \ \ \ \ \ \isacommand{proof}\ (rule\ Weierstrass\_m\_test)\isanewline
\ \ \ \ \ \ \ \ \isacommand{let}\ ?M\ =\ "\isasymlambda n.\ (1\ +\ norm\ z)\ \isacharcircum \ n\ /\ fact\ n"\isanewline
\ \ \ \ \ \ \ \ \ \ \isacommand{show}\ "summable\ ?M"\isanewline
\ \ \ \ \ \ \ \ \ \ \ \ \prf\isanewline
\ \ \ \ \ \ \ \ \isacommand{fix}\ n\ z'\isanewline
\ \ \ \ \ \ \ \ \isacommand{assume}\ \ "z'\ \isasymin \ cball\ z\ 1"\isanewline
\ \ \ \ \ \ \ \ \isacommand{show}\ "norm\ (coeff\ n\ *\ p\ n\ z')\ \isasymle \ ?M\ n"\isanewline
\ \ \ \ \ \ \ \ \ \ \prf\isanewline
\ \ \ \ \ \ \isacommand{qed}
\end{isabelle}
The hard part is over. Checking that the image of $\{\zeta_\alpha : \alpha < \gamma\}$ under \isa{hh} lies within $D$ is easy, because for every complex number of the form $w_n$, the summation is actually finite.
\begin{isabelle}
\ \ \ \ \ \ \isacommand{have}\ hh\_eq\_dd:\ "hh\ (w\ n)\ =\ dd\ n\ coeff"\ \isakeyword{for}\ n\isanewline
\ \ \ \ \ \ \ \ \prf\isanewline
\ \ \ \ \ \ \isacommand{then}\ \isacommand{have}\ "hh\ (w\ n)\ \isasymin \ D"\ \isakeyword{for}\ n\isanewline
\ \ \ \ \ \ \ \ \isacommand{using}\ DD\_def\ dd\_in\_DD\ \isacommand{by}\ fastforce
\end{isabelle}

Last comes the diagonalisation argument, showing that the function just constructed is indeed new:
\begin{isabelle}
\ \ \ \ \ \ \isacommand{have}\ "hh\ (w\ n)\ \isasymnoteq \ f\ (\isasymeta \ n)\ (w\ n)"\ \isakeyword{for}\ n\isanewline
\ \ \ \ \ \ \ \ \isacommand{using}\ DD\_def\ dd\_in\_DD\ g\_def\ hh\_eq\_dd\ \isacommand{by}\ auto\isanewline
\ \ \ \ \ \ \isacommand{then}\ \isacommand{show}\ "\isasymforall \isasymbeta \isasymin elts\ \isasymgamma .\ hh\ \isasymnoteq \ f\ \isasymbeta "\isanewline
\ \ \ \ \ \ \ \ \isacommand{by}\ (metis\ \isasymeta \ bij\_betw\_imp\_surj\_on\ imageE)
\end{isabelle}

\subsection{Concluding the Proof}

About 20 lines of boilerplate are needed to code up the formal application of transfinite recursion to the construction formalised above. Skipping over this, let's look at the conclusion of the proof.
We finally have a family of analytic functions $\{f_\beta : \beta < \omega_1 \}$ satisfying~(\ref{eqn:f_in_D}) above. 

We now show that for all $z$, the set $\{f_\beta(z) : \beta < \omega_1 \}$ is countable by writing $z=\zeta_\alpha$ for some $\alpha<\gamma$.
The point is that 
\begin{align*}
\{f_\beta(\zeta_\alpha) : \beta < \omega_1 \} = \{f_\beta (\zeta_\alpha): \beta\le\alpha\} \cup \{f_\beta (\zeta_\alpha): \alpha<\beta\}
\end{align*}
and that both parts of the union are countable: the first is a collection below~$\alpha$, a countable ordinal, and the second is a subset of~$D$, a countable set.
In the proof below, \isa{?B} is the set $\{\beta: \alpha<\beta<\gamma\}$, while \isa{elts\ (succ\ \isasymalpha)} is the set of all ordinals up to~$\alpha$. 

\begin{isabelle}
\ \ \isacommand{show}\ ?thesis\isanewline
\ \ \isacommand{proof}\isanewline
\ \ \ \ \isacommand{let}\ ?F\ =\ "f\ `\ elts\ \isasymomega 1"\isanewline
\ \ \ \ \isacommand{have}\ "countable\ ((\isasymlambda f.\ f\ z)\ `\ f\ `\ elts\ \isasymomega 1)"\ \isakeyword{for}\ z\isanewline
\ \ \ \ \isacommand{proof}\ -\isanewline
\ \ \ \ \ \ \isacommand{obtain}\ \isasymalpha \ \isakeyword{where}\ \isasymalpha :\ "\isasymzeta \ \isasymalpha \ =\ z"\ "\isasymalpha \ \isasymin \ elts\ \isasymomega 1"\ "Ord\ \isasymalpha "\isanewline
\ \ \ \ \ \ \ \ \isacommand{by}\ (meson\ Ord\_\isasymomega 1\ Ord\_in\_Ord\ UNIV\_I\ \isasymzeta \ bij\_betw\_iff\_bijections)\isanewline
\ \ \ \ \ \ \isacommand{let}\ ?B\ =\ "elts\ \isasymomega 1\ -\ elts\ (succ\ \isasymalpha )"\isanewline
\ \ \ \ \ \ \isacommand{have}\ eq:\ "elts\ \isasymomega 1\ =\ elts\ (succ\ \isasymalpha)\ \isasymunion \ ?B"\isanewline
\ \ \ \ \ \ \ \ \isacommand{using}\ \isasymalpha \ \isacommand{by}\ (metis\ Diff\_partition\ Ord\_\isasymomega 1\ OrdmemD\ less\_eq\_V\_def\ succ\_le\_iff)\isanewline
\ \ \ \ \ \ \isacommand{have}\ "(\isasymlambda f.\ f\ z)\ `\ f\ `\ ?B\ \isasymsubseteq \ D"\isanewline
\ \ \ \ \ \ \ \ \isacommand{using}\ \isasymalpha \ inD\ \isacommand{by}\ clarsimp\ (meson\ Ord\_\isasymomega 1\ Ord\_in\_Ord\ Ord\_linear)\isanewline
\ \ \ \ \ \ \isacommand{then}\ \isacommand{have}\ "countable\ ((\isasymlambda f.\ f\ z)\ `\ f\ `\ ?B)"\isanewline
\ \ \ \ \ \ \ \ \isacommand{by}\ (meson\ \isacartoucheopen countable\ D\isacartoucheclose \ countable\_subset)\isanewline
\ \ \ \ \ \ \isacommand{moreover}\ \isacommand{have}\ "countable\ ((\isasymlambda f.\ f\ z)\ `\ f\ `\ elts\ (succ\ \isasymalpha ))"\isanewline
\ \ \ \ \ \ \ \ \isacommand{by}\ (simp\ add:\ \isasymalpha \ less\_\isasymomega 1\_imp\_countable)\isanewline
\ \ \ \ \ \ \isacommand{ultimately}\ \isacommand{show}\ ?thesis\isanewline
\ \ \ \ \ \ \ \ \isacommand{using}\ eq\ \isacommand{by}\ (metis\ countable\_Un\_iff\ image\_Un)\isanewline
\ \ \ \ \isacommand{qed}\isanewline
\ \ \ \ \isacommand{then}\ \isacommand{show}\ "Wetzel\ ?F"\isanewline
\ \ \ \ \ \ \isacommand{unfolding}\ Wetzel\_def\ \isacommand{by}\ (blast\ intro:\ anf)\isanewline
\ \ \ \ \isacommand{show}\ "uncountable\ ?F"\isanewline
\ \ \ \ \ \ \isacommand{using}\ Ord\_\isasymomega 1\ countable\_iff\_less\_\isasymomega 1\ countable\_image\_inj\_eq\ injf\ \isacommand{by}\ blast\isanewline
\ \ \isacommand{qed}\isanewline
\isacommand{qed}
\end{isabelle}

The following corollary to the two cases summarises the equivalence between the Wetzel property and the negation of CH:

\begin{isabelle}
\isacommand{theorem}\ Erdos\_Wetzel:\ "C\_continuum\ =\ \isasymaleph 1\ \isasymlongleftrightarrow \ (\isasymexists F.\ Wetzel\ F\ \isasymand \ uncountable\ F)"\isanewline
\ \ \isacommand{by}\ (metis\ C\_continuum\_ge\ Erdos\_Wetzel\_CH\ Erdos\_Wetzel\_nonCH\ less\_V\_def)
\end{isabelle}

\section{Discussion}

The formalisation of mathematics within proof assistants is being pursued actively. There are too many contributions to list, but notable recent ones include the Buzzard--Commelin--Massot formalisation of perfectoid spaces~\cite{buzzard-perfectoid} and the striking progress on the Liquid Tensor Experiment, which is led by Johan Commelin.%
\footnote{\url{https://tinyurl.com/5n8rh297}}
Both of these involve formalising the sophisticated work of Fields Medallist Peter Scholze, using Lean.

The small example reported here cannot be compared with such accomplishments, but shines a light on an issue that makes mathematics difficult: its interconnectedness. The connections between number theory and complex analysis are well known. If your problem concerns itself with sets of permutations, you might suddenly find yourself needing group theory. If you are trying to count the elements of the group, you might find yourself requiring sophisticated combinatorial arguments. If you find yourself talking about countable sets (and countability is an everyday notion), you could find yourself wandering into the world of cardinals, then into the world of ordinals and then perhaps into the undecidable. To quote Wetzel himself, ``once again a natural analysis question has grown horns''~\cite[p.\ts244]{garcia-wetzels-problem}.

I created the ZFC-in-HOL library in October 2019 with no specific application in mind but merely thinking that it might be useful occasionally to talk about really big sets.
Then I decided to apply it to a project that had been proposed by Mirna Džamonja and Angeliki Koutsoukou-Argyraki: to formalise some ordinal partition theory. This concerns advanced generalisations of Ramsey's theorem; the formalisation~\cite{Ordinal_Partitions-AFP} was complete by August 2020 and a paper is now available~\cite{dzamonja-formalising}.
However, this work essentially belongs to straight set theory, borrowing little from Isabelle/HOL apart from natural numbers and lists.

The Wetzel example showed that there were a few gaps limiting the integration between that theory and Isabelle/HOL, above all, a more general definition of cardinality. It's easy to define the cardinality of an arbitrary Isabelle/HOL set (i.e. having a type of the form \isa{T~set} as the least ordinal whose elements can be put into bijection with that set.
The equivalent notion of cardinality for ZF sets is a trivial consequence.

The original AFP entry~\cite{Wetzels_Problem-AFP}, published online in February 2022, represents about three weeks' work.
It begins with a variety of extensions to the Isabelle/HOL libraries, showing precisely how the original ZFC library needed to be extended.
It includes an early and rather cumbersome definition of cardinality for HOL sets. In updated versions of the development,%
\footnote{At \url{https://devel.isa-afp.org/entries/Wetzels_Problem.html}. Future readers should be able to locate the original Wetzel entry via ``Older releases'' on the download page.}
the library material has been moved elsewhere. The latest version is 360 lines, 2833 tokens, which is not bad compared with the original text by Erdős~\cite{erdos-interpolation}, which is one and a half pages long. The exposition by Aigner and Ziegler~\cite{aigner-proofs} is two full pages long (67 lines, 1026 tokens). A crude comparison based on compressing the texts by gzip suggests a de Bruijn factor of under~2.3. For the original version, including the material later moved to libraries, the de Bruijn factor is 4.2.

\section{Conclusions}

The intriguing solution to Wetzel's problem by Erd\H{o}s was easily formalised in Isabelle/HOL, using a library designed to integrate set theory with higher-order logic.
The ease with which we can intermix analytic functions with ordinals gives reason to hope that harder mixed-domain problems will be amenable to formalisation without particular difficulties.
It would be interesting to see this example tackled using other proof assistants.

\subsubsection{Acknowledgements} 
This work was supported by the ERC Advanced Grant ALEXANDRIA (Project GA 742178). 
Dmitriy Traytel provided a particularly slick formal axiomatisation of type~$V$. 
Manuel Eberl provided a crucial tip for the CH case.
Angeliki Koutsoukou-Argyraki and the referees made insightful suggestions.

For the purpose of open access, the author has applied a Creative Commons Attribution (CC BY) licence to any Author Accepted Manuscript version arising from this submission.

\bibliographystyle{splncs04}
\bibliography{string,atp,general,isabelle,theory,crossref}

\end{document}